\documentclass[reprint,amsmath,amssymb,aps,prb,superscriptaddress]{revtex4-1}
\usepackage{physics}
\usepackage{units}
\usepackage{amsmath}
\usepackage{url}
\usepackage{natbib}
\usepackage{textcase}
\usepackage{amssymb}
\usepackage{graphicx}
\usepackage{bm}
\usepackage{float}
\usepackage{multirow,microtype,color,relsize,ulem}
\usepackage{braket}
\usepackage{subfigure}
\usepackage[breaklinks=true]{hyperref}
\usepackage[utf8]{inputenc}
\usepackage[english]{babel}
\hypersetup{colorlinks=true,linkcolor=blue,citecolor=blue}
\hypersetup{linktocpage}
\usepackage{CJK}
\usepackage{breakcites}
\usepackage{float}

\begin{document}
\begin{CJK*}{UTF8}{gbsn}
\title{Anderson transition in three-dimensional systems with non-Hermitian disorder}

\author{Yi Huang~(黄奕)}
\email[Corresponding author: ]{huan1756@umn.edu}
\author{B. I. Shklovskii}
\affiliation{School of Physics and Astronomy, University of Minnesota, Minneapolis, MN 55455, USA}

\begin{abstract}	
We study the Anderson transition for three-dimensional (3D) $N \times N \times N$ tightly bound cubic lattices where both real and imaginary parts of onsite energies are independent random variables distributed uniformly between $-W/2$ and $W/2$. 
Such a non-Hermitian analog of the Anderson model is used to describe random-laser medium with local loss and amplification. 
We employ eigenvalue statistics to search for the Anderson transition. 
For 25\% smallest-modulus complex eigenvalues we find the average ratio $r$ of distances to the first and the second nearest neighbor as a function of $W$. 
For a given $N$ the function $r(W)$ crosses from $0.72$ to 2/3 with a growing $W$ demonstrating a transition from delocalized to localized states. 
When plotted at different $N$ all $r(W)$ cross at $W_c = 6.0 \pm 0.1$ (in units of nearest neighbor overlap integral) clearly demonstrating the 3D Anderson transition. 
We find that in the non-Hermitian 2D Anderson model, the transition is replaced by a crossover.    
\end{abstract}

\date{\today}

\maketitle
\end{CJK*}

\par Anderson localization is the central concept of solid state physics for more than 60 years~\cite{Andersonoriginal,Anderson2D,Anderson1972}. It determines electron conductivity of doped crystalline and amorphous semiconductors and many other disordered systems and is observed in experiments~\cite{Mottbook,EfrosShklovskii1984}.  

In recent years the problem of localization attracted renewed interest as research moved to formerly unexplored area of non-Hermitian systems. 
Random lasers~\cite{rl,Wiersma:2013aa,rl2,rkt} with random dissipation and amplification regions are such prototypical non-Hermitian systems. 
The other parts of non-Hermitian disorder physics are related to Hatano-Nelson matrices~\cite{hn}, their biological applications~\cite{Amir2016,Nelson2019} or to spin chains~\cite{Posen2019, Hamazaki2019a, Hamazaki2019}.
All these works focus on one-dimensional systems.

A simple and elegant extension of the 2D Anderson localization problem was proposed in a recent paper by Tzortzakakis, Makris and Economou (TME)~\cite{Economou2019}. 
They studied 50 $\times$ 50 tight-binding square lattices with real overlap energy $I_{ij} = I$, and random complex onsite energies $E_i$ whose real and imaginary parts are independent random variables distributed uniformly between $-W/2$ and $W/2$.
The Hamiltonian reads
\begin{equation}\label{eq:h}
	H = \sum_i E_i a^{\dagger}_i a_i - \sum_{i,j} (I_{ij} a^{\dagger}_i a_j + \text{h.c.}),
\end{equation}
where $i,j$ in the second term are nearest neighbors, and the hard-wall boundary is employed (no bonds extended out the boundary).
Below we call this non-Hermitian Hamiltonian the TME model. 
By calculating the participation ratio of eigenfunctions of such non-Hermitian Hamiltonian, TME noticed that they become progressively more localized when $W$ (in units of $I$) grows from 1 to 5. 
Simultaneously the distribution function $P(s)$ of nearest neighbor distances $s$ between eigenvalues in the complex plane widens, which shows that the repulsion of eigenvalues weakens due to the progressive localization of eigenfunctions. 
This behavior is similar to what happens in the Anderson model~\cite{Shklovskii1993}. 
TME, however, did not raise a question whether there is an Anderson transition or a crossover in the limit of large system. 

In this paper, we focus on the question of the existence of the Anderson transition in the TME model for 3D cubic and 2D square lattices. 
We show that in the TME model, the Anderson transition exists in 3D, but is missing in 2D, as in the conventional Anderson model~\cite{Anderson2D,Shklovskii1993}. 

To identify the Anderson transition in 3D we follow Ref.~\onlinecite{Shklovskii1993} and use statistics of complex eigenvalues obtained by diagonalization of the TME model on many realizations of $N \times N \times N$ cubic lattices.
The diagonalization is done using LAPACK~\cite{LAPACK}.
We do this for $N$=8, 10, 12, 16, and 20 at $W$=4, 5, 5.5, 6, 6.5, 7, and 8. 
For analysis of the spectrum we need a parameter uniquely characterizing the statistics of eigenvalues at a given $W$ and $N$. 
For the Anderson model originally this parameter was an area of the large $s$ tail of nearest neighbor distribution function $P(s)$~\cite{Shklovskii1993}, but later Ref.~\onlinecite{Oganesyan2007} suggested a better measure $r = \ev{\min(s_{i-1}/s_{i}, s_{i}/s_{i-1})}$, where $s_i$ is the spacing between $i$-th and $(i+1)$-st energy levels, and $\ev{\dots}$ stands for the average over the studied part of the spectrum and over realizations. 
For the TME model, where eigenvalues are points in the complex plane (see, for example, Ref.~\onlinecite{Economou2019}) we have chosen the parameter $r(W) = \ev{s_1/s_2}$, where $s_1$ and $s_2$ are distances from a given eigenvalue to its first and second nearest eigenvalues.~\footnote{Our $r(W)$ defined by ratio of spacings to first and second nearest eigenvalues is actually different from the $r$ introduced by Oganesyan and Huse~\cite{Oganesyan2007} 
when applied to real spectrum of the Anderson model. Contrary to the definition of Oganesyan and Huse, in our case we allow the first and second nearest level to be on the same side of $\epsilon_i$. 
We checked that both definitions of $r(W)$ work for the conventional Anderson model and lead to the Anderson transition at the same $W=16$.}
Our parameter $r(W)$ is the modulus of the more informative complex parameter introduced in Ref.~\onlinecite{Posen2019}.
Similar to Ref.~\onlinecite{Economou2019} we found that eigenvalues near the rectangular border of the complex spectrum correspond to more localized states. 
Therefore, to deal with eigenvalues with similar localization properties we calculated $r$ of 25\% smallest-modulus eigenvalues selected by a rectangular window whose sides are roughly twice smaller than the whole spectrum.
The number of random realizations varied with $N$ in such a way that the number of studied eigenvalues at each combination of $W$ and $N$ was kept around $2\times 10^5$.

\begin{figure}[t]
	\centering
	\includegraphics[width=\linewidth]{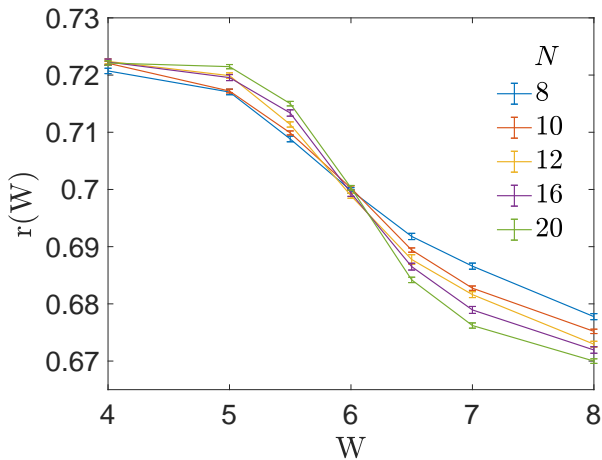}
	\caption{The ratio $r(W)$ in the 3D cubic lattice at different $N$. The crossing in the middle indicates the Anderson transition at $W_c = 6.0 \pm 0.1$.}
	\label{fig:3d}
\end{figure}

\begin{figure}[t]
	\centering
	\includegraphics[width=\linewidth]{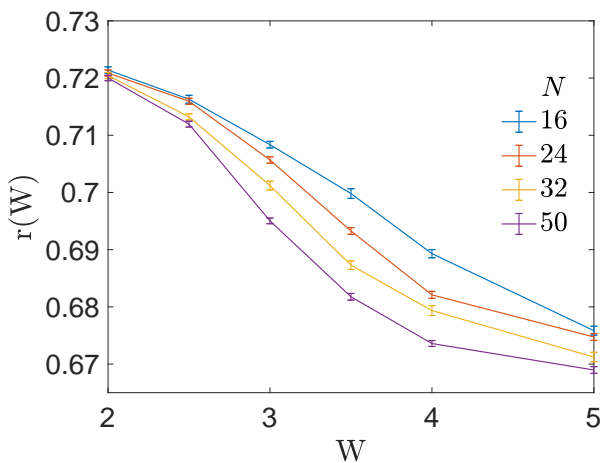}
	\caption{The ratio $r(W)$ in the 2D square lattice at different $N$. These curves do not cross showing that there is no Anderson transition.}
	\label{fig:2d}
\end{figure}

Fig.~\ref{fig:3d} shows our results for $r(W)$ plotted as a function of $W$ at different $N$. 
We see that all curves $r(W)$ with growing $W$ cross over from the ``Wigner surmise value'' 0.72 to the Poisson value 2/3 calculated for random points in a plane in Ref.~\onlinecite{Posen2019}. 
Remarkably, all curves $r(W)$ cross each other near $W_c = 6$. 
This means that in the limit of large $N$ there is an Anderson transition at $W_c = 6.0 \pm{0.1}$ for TME model.
This transition point is much smaller than $W_c = 16$ of the conventional Anderson model~\cite{Zharekeshev2001}.
Apparently, non-Hermitian disorder is more effective for the localization of wavefunctions. 
We believe this effectiveness results from larger absolute values of locator expansion energy denominators~\cite{Anderson1972}, particularly for small-modulus energy eigenvalues. 
\begin{figure}[t]
	\centering
	\includegraphics[width=\linewidth]{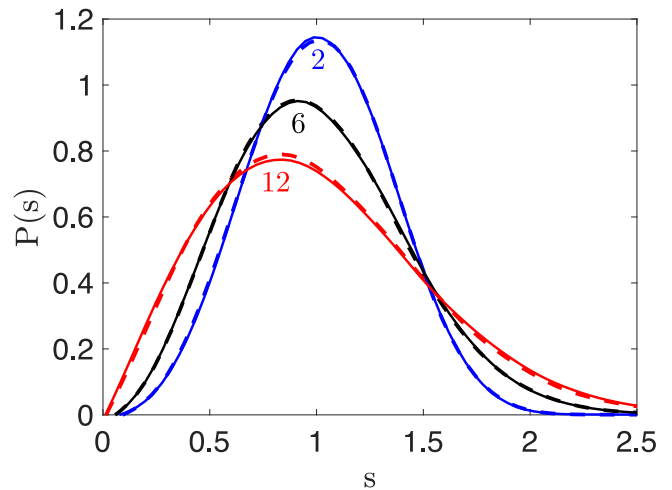}
	\caption{The probability density $P(s)$ as a function of (unfolded) spacing between complex eigenvalues $s$. Dashed and continuous curves correspond to 3D TME model with $N$=8 and 16 respectively. Disorder width $W$ is shown by numbers next to each color pair of curves.}
	\label{fig:ps}
\end{figure}

As we mentioned above our statistical analysis used a window located around the origin in the complex plane containing 1/4 of all eigenvalues. 
We checked at $N=12$ that when we shrink this window to the fraction 1/16 (and compensate for the loss of statistical samples by increasing of the number of realizations correspondingly), the $r(W)$ curve near $W=W_c$ shifts to larger $W$ by 0.15. 
Further window shrinking to $1/64$ does not change $r(W)$.
Thus, our estimate for $W_c$ in the limit of the shrinking window is $6.15\pm{0.15}$.

From Fig.~\ref{fig:3d}, we also analyze the critical scaling behavior of $r(W,N)$ near $W_c$ as a function of $(W - W_c)N^{1/\nu}$, and find the critical exponent $\nu = 1.5 \pm 0.2$ similar to Ref.~\onlinecite{Shklovskii1993}.

To emphasize the non-trivial nature of the 3D Anderson transition seen in Fig.~\ref{fig:3d} we present in Fig.~\ref{fig:2d} our results of the similar study for 2D square lattices. 
We see that while all curves $r(W)$ still cross over from 0.72 to 0.67, they obviously avoid intersections with each other.
This means that in the 2D TME model there is no Anderson transition (like in the conventional 2D Anderson model).
Qualitatively similar behavior of the eigenvalue statistics with increasing disorder was studied in 1D non-Hermitian systems~\cite{Nelson2019}.

Now we return to the 3D TME model.
Fig.~\ref{fig:ps} presents the probability density $P(s)$ of spacing $s$ to the nearest neighbor eigenvalue in the complex energy plane for two different $N$ and three values of $W$. 
The black curves correspond to critical point $W=6$ of  the found above TME model Anderson transition and two other values of $W$ are chosen to be far from the transition on both sides of it. 
To eliminate the role of changing density of states (unfolding the spectrum) we evaluate the level spacing $s$ in units of local average level spacing $\ev{s}$ calculated in a 100$\times$100 mesh. 

The most important result seen in Fig.~\ref{fig:ps} is that black curves for substantially different $N = 8$ and $N = 16$ are identical. This confirms that at $W=6$ the size effect is absent and that $W=6$ is indeed the Anderson transition point. 
This also confirms the validity of our method using a single parameter $r(W)$ to characterize $P(s)$. 
Thus, the black line $P(s)$ represents the new universal transition point statistics in complex plane similar to the one discovered earlier for the Anderson model for energy level spacing~\cite{Shklovskii1993}. 
The red curves are also very close to each other and to the asymptotic at large $N$ 2D Poisson distribution $P(s) = \pi s/2 \ {e}^{-(\pi/4)s^2}$~\cite{Haake1991}.
The blue curves are very close to each other and to the aymptotic at large $N$ ``Wigner-Dyson'' distribution for TME model, which belongs to the universality class AI$^{\dagger}$~\cite{Hamazaki2019}. (Both asymptotic are not shown here). 
This happens because we intentionally have chosen $W$ for red and blue curves to be far from the transition. 
Closer to transition size effects are more obvious as seen from Fig.~\ref{fig:3d}. 
Thus, Fig.~\ref{fig:ps} practically shows all three universal statistics of 3D TME model. 

We also explored another non-Hermitian model different from TME trying to see how general 3D Anderson transition is.
For this model, the diagonal matrix elements are the same as in the TME model, while the overlap energy is $I_{ij} = -I_{ji}$, and $I_{ij}$ is a random variable with 50\% probability to be $\pm 1$.
We find the 3D Anderson transition at $W_c = 6.15\pm 0.15$ with critical exponent $\nu = 1.5 \pm 0.2$, same as for TME.

\begin{acknowledgements}
We thank A. Kamenev and D. R. Nelson for discussions, and H. Wang and Y. Yang for technical support.
Calculations by Y.H. were supported primarily by the NSF through the University of Minnesota MRSEC under Award No. DMR-1420013.
The authors acknowledge the Minnesota Supercomputing Institute (MSI) at the University of Minnesota for providing resources that contributed to the research results reported within this paper.
\end{acknowledgements}


%

\end{document}